\documentclass[12pt,letterpaper,english]{article}
\usepackage{babel}
\usepackage{graphics}
\usepackage[T1]{fontenc}
\begin{document}

\centerline {\Large An Electrical Network Model of Plant Intelligence$^*$}

\bigskip

\bigskip

\bigskip

\centerline {Bikas K. Chakrabarti$^{(1)}$ and Omjyoti Dutta$^{(2)}$}

\centerline {$^{(1)}$Saha Institute of Nuclear Physics, 1/AF Bidhan Nagar, 
Kolkata-700064.}
\centerline {$^{(2)}$Electrical Engineering Dept. (4th Yr.), 
Jadavpur University, Kolkata-700032.}

\vskip 3 cm

\centerline {\bf Abstract }

\medskip

\noindent A simple electrical network model, having
logical gate capacities, is proposed here for computations 
in plant cells. It is compared and contrasted with the animal brain network
structure and functions.

\vskip 6 cm

\noindent------------------------------------------------

\noindent $^*$ Talk at the Condensed Matter Days-2002, held in Bhagalpur
University, Bhagalpur, during August 29-31, 2002 (to be published in Ind.
J. Phys.).

\newpage

\noindent Considerable investigations and efforts have been made in 
understanding how the animal and human brains compute and recognise 
various spatial and temporal patterns [1,2]. The essential model consists of
a network of large number (about 10$^{12}$ in case of human, 10$^{6}$ in 
case of birds) of two state electrical devices  called neurons which are
capable of just summing over the various  (milli-volt order) input 
electrical pulses for  a short synaptic period (of milli-second order) 
collected by the (10$^6$ or so) dendrites of each neuron, and comparing this
sum with a threshold. The synaptic interactions among the neurons develop
during the ``learning" process, and can 
be both excitory or inhibitory, rendering
the network randomly frustrated. The computational capabilities emerge out
of the collctive dynamics of the network, which is nonlinear (due to the
threshold behaviour of each neuron). For symmetric interactions, one can
define an energy function (or free energy at finite noise or ``temperature"
level) for the network and the local free energy minima corresponds to 
the various local attractor patterns or memory states of the network
(Hopfield [1]). For long-range interactions, the statistical physics of
such a network is analytically tractable to a large extent (Amit et al
[1, 2]). The processing of informations in such network models and their
detailed analysis are now established (see e.g., Nishimori [2]). These
demonostrated capabilities of such networks are of course very limited 
in their emerging computational abilities [2] and far short of anything
like consciousness, where some aspects of quantum mechanics (entanglement
in the molecules in microtubules of a single neuron) are speculated to be
involved [3].

Are the plants around us intelligent? Do they also deserve our attention 
in this context of modelling for information processing and computation? 
Plants have remarkable adaptibility in changed environments. They 
survive in every landscape of this earth, representing almost 99$\%$ of 
its biomass. Such marvellous adaptive behaviour must be interpreted
to be intelligent; although naive definitation of intelligence seem to
involve movement of the animal (either bodily or part of it) and plants
can not move (bodily) [4]. How such intelligent behaviour of mindless
plants, having no brain, compare with those of animals [5]? Plants do not
have neuronal cells either.

Almost eighty years back, Bose detected electrical signalling between 
plant cells in coordinating its responses to the environment [6,7]. 
Although the chemical diffusion of (uncharged) molecules 
 is a dominent source of signalling between the plant cells, it is a
very slow mode. It is now established [7] that some signals are 
trasmitted within the plants at much smaller time scale (with signal
velocity about 30-400 mm/sec, depending on the plant and its environment).
Such fast transmissions are due to electrical pulses, generated by ionic 
motions within the plant cells. Although not the dominent mode, except in 
some very sensetive plants like Desmodium or Mimosa [7], the electrical
mode (due to migration of Ca$^+$, K$^+$, etc ions)  generally present in
the cells of  all
the plants [8]. However, these electrically excitable plant cells do not
have many dendrites, like for the neurons, nor are they connected by random 
excitory/inhibitory (frustrating) interactions. 

In absence of the highly connected (frustrated) network of neuron-like 
units, as in the animal brains, the plants might be utilising the non-linear
current (I)-voltage (V) characterestics of their cell membranes  for the
logical operations (gates). In fact, the  plant vacuolar membrane 
current-voltage characterestics [9] is now established to
be equivalent to that of a Zener diode,
as indicated in Fig. 1.

\vskip 0.4in
\vspace{0.3cm}
{\centering \resizebox*{9cm}{6cm}{\rotatebox{0}{\includegraphics{fig1.eps}}} \par}
\vspace{0.3cm}
\vskip 0.4in

%\begin{figure}
%\centerline{\psfig{file=fig1.eps,height=6cm,width=8cm}}
%\end{figure}

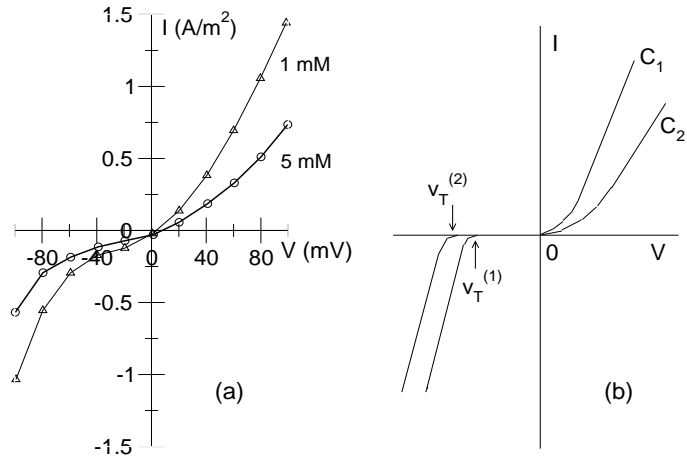
\noindent {\small Fig. 1. (a) Typical I-V curves of the plant vacuolar 
membrane fast-activating channel (from [9]). The current being mainly due to
Ca$^+$ ions and the (reversible) effect of divalant Putrescine 
(C$_4$H$_{14}$N$_2^{2^+}$) are shown at 
two different concentrations. (b) The 
equivalent Zener diode-like behaviour of the membrane, where the Putrescine 
concentrations ($c$) modulate the changes in the threshold voltage V$_T$.}

\medskip

One can utilise such a threshold behaviour of the plant cell membranes
to develop or model gates for perfoming simple logical operations. In
Fig. 2, such a model network containing four such threshold units; one in
the output and the other three in the input. 
Each of these threshold units is modelled as a binary unit, having two states:
0 and 1. The inter-unit connection strength is denoted here by the matrix
$W$. The output $O$ of the network considerd can then be expressed
as

$$ O = \theta (I - \phi), \eqno (1) $$

\noindent where $\theta$ is the step function ($\theta(x) = 1$ for
$x \ge 0$; and 0 otherwise),
 
$$ I = W_1 I_1 + W_2I_2+W_3 I_3, \eqno (1a)$$

\noindent and $\phi$ is the threshold strength (determined by the threshold
voltage $V_T$) for the output unit.  $I_1$, $I_2,$ $I_3$ are the 
inputs to the three input units and $W_1$, $W_2$, $W_3$ are their 
connectivity strengths with the output unit, as indicated in Fig. 2.

\vskip 0.2in

\vspace{0.3cm}
{\centering \resizebox*{7cm}{5cm}{\rotatebox{0}{\includegraphics{fig2.eps}}} \par}
\vspace{0.3cm}
\vskip 0.2in

%\begin{figure}
%\centerline{\psfig{file=fig2.ps,height=6cm,width=9cm}}
%\end{figure}

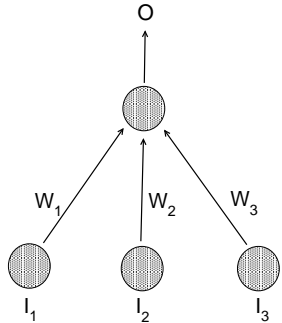
\noindent {\small Fig. 2. A simple network containing four threshold 
units (three in the input and one in the output) for performing logical 
operations like AND, OR, NAND, etc.}

\medskip

For different combinations of $I_1$, $I_2$ and $I_3$, the outputs for
different logic gates are given in the Table 1. These can be easily achieved 
using  the combinations of inter-cell connections and the output
cell thresholds as:

$$ W_1 = W_2 = W_3 = 1, \phi = 3, \eqno (2a)$$

\noindent for the AND gate;

$$ W_1 = W_2 = W_3 = 1, \phi = 1, \eqno (2b)$$

\noindent for the OR gate; and

$$ W_1 = W_2 = W_3 = -1, \phi = -2 \eqno (2c)$$

\noindent for the NAND gate.

\medskip 
\vskip 0.2in

\centerline {Table I: The input-output (truth) table for the logic gates}
\begin{center}
\begin{tabular}{|ccc|ccc|}
\hline
{$~$} & {Inputs} & {$~$} & {$~$} & {Output O} & {$~$}\\ \hline
{$I_1$} & {$I_2$} & {$I_3$} & {AND} & {OR} & {NAND}\\ \hline
{0} & {0} & {0} & {0} & {0} & {1}\\ \hline
{0} & {0} & {1} & {0} & {1} & {1}\\ \hline
{0} & {1} & {0} & {0} & {1} & {1}\\ \hline
{0} & {1} & {1} & {0} & {1} & {1}\\ \hline
{1} & {0} & {0} & {0} & {1} & {1}\\ \hline
{1} & {0} & {1} & {0} & {1} & {1}\\ \hline
{1} & {1} & {0} & {0} & {1} & {1}\\ \hline
{1} & {1} & {1} & {1} & {1} & {0}\\ \hline
\end{tabular}
\end {center}
\vskip 0.2in

These gate capabilities of simple networks of the plant cell membranes,
using their nonlinear characterestics and cosequent threshold behaviour 
(with adjustable thresholds through changed concentrations of, for example,
the putrescine and the interaction matrix {\bf $W$}) would allow (cf. [2])
simple computations in the electrical channels of the plants. It may be noted
that such networks here are much more local and tiny in structure, compared 
to the massively connected and parallelly working network of animal
brains. Also, the network matrix {\bf $W$} elements are either all positive
(excitory) or all negative (inhibitory) in eqns. (2). As such, they 
do not involve any frustration as in the animal brains and have got
consequently several limitations in their computational capabilities; for
example, they lack the distributed parallel computational capacity, associate
memory, etc.

\vskip 1 cm

We are grateful to Indrani Bose, Arnab Chatterjee and Dibyendu Sengupta
 for many useful comments and suggestions.

\vskip 2 cm

\noindent {\bf References:}

\medskip

\noindent [1] J. J. Hopfield, ``Neural 
Networks and Physical Systems with Emergent
Computational Abilities'', Proc. Nat. Acad. Sc. (USA) {\bf 79} 2554 (1982);
D. J. Amit, H. Gutfreund and Sompolinsky, ``Statistical Mechanics of 
Neural Networks Near Saturation'', Ann. Phys. {\bf 173} 30 (1987).

\medskip

\noindent [2] For reviews see e.g., J. Hertz, A. Krough and R. G. Palmer,
{\it Introduction to the Theory of Neural Computation}, Addison-Wesley,
Reading, MA (1991); ~~ B. K. Chakrabarti and P. K. Dasgupta, ``Modelling Neural
Networks'', 
 Physica {\bf 186} 33 (1992); H. Nishimori, {\it Statistical
Physics of Spin Glasses and Information Processing}, Oxford Univ.
Press, Oxford (2001).

\medskip

\noindent [3] R. Penrose, {\it The Emperor's New Mind}, Oxford Univ. Press,
Oxford (1999); M. A. Nielsen and I. L. Chuang, {\it Quantum Computation and
Quantum Information}, Cambridge Univ. Press, Cambridge (2000).

\medskip

\noindent [4] A. Trewavas, ``Mindless Mastery'', Nature {\bf 415} 841 (2002).

\medskip

\noindent [5]  B. K. Chakrabarti, ``Mathematics, Brain Modelling and Indian
Concept  of Mind'', Proc. Asiatic Soc., Kolkata (in press); also at 
arXiv:cond-mat/0205094 (2002).

\medskip

\noindent [6] J. C. Bose, {\it The Nervous Mechanism of Plants}, Longmans, 
Green \& Co, London (1923).

\medskip

\noindent [7] V. A. Shepherd, ``Bioelectricity and Rhythms of Sensitive  
Plants: The Biophysical Research of J. C. Bose'', 
Current Sc. {\bf 77} 189 (1999).

\medskip

\noindent [8] T. Genoud and J.-P. M\`etraux, ``Crosstalk in Plant Signalling'', 
Trends in Plant Sc., {\bf 4} 503 (1999).

\medskip

\noindent [9] L. I. Br\"uggemann, I. I. Pottosin and G. Schonknecht, 
``Cytoplasmic Polyamines Block Fast-Activating Vacuolar Cation Channel'',
The Plant J. {\bf 16} 101 (1998). 

\end{document}